\documentstyle[12pt,amssymb]{article}
\def\dalemb#1#2{{\vbox{\hrule height .#2pt
        \hbox{\vrule width.#2pt height#1pt \kern#1pt
                \vrule width.#2pt}
        \hrule height.#2pt}}}

\newcommand{\be}{\begin{equation}}
\newcommand{\ee}{\end{equation}}
\newcommand{\ba}{\begin{array}}
\newcommand{\ea}{\end{array}}
\newcommand{\bea}{\begin{eqnarray}}
\newcommand{\eea}{\end{eqnarray}}
\def\nn{\nonumber}
\def\ft#1#2{{\textstyle{{\scriptstyle #1}\over {\scriptstyle #2}}}}

\def\sst#1{{\scriptscriptstyle #1}}
\def\st#1{{\scriptstyle #1}}

\def\Z{{\Bbb Z}}
\def\im{{\rm i}}
\def\ie{{\it i.e.\ }}
\def\eg{{\it e.g.\ }}

\begin{document}
\begin{flushright}
\hfill{Imperial/TP/1-02/24}\\
\hfill{hep-th/0208055}
\end{flushright}

\thispagestyle{empty}

\begin{center}

{\Large\bf Domain Walls and Spaces of Special Holonomy}\\
\vspace{.5cm}

{\sc K.S. Stelle\footnote{Research supported in part by the European Community's
Human Potential Programme under contract HPRN-CT-2000-00131 Quantum Spacetime,
and by PPARC under SPG grant No. 613}}\\

\vspace{.5cm}
{\em The Blackett Laboratory, Imperial College} \\
{\em London SW7 2BZ, UK} \\

\vspace{.5cm}
\centerline{\bf Abstract}
\end{center}
\begin{quote}\small
We review the relations between a family of domain-wall solutions to M-theory
and gravitational instantons with special holonomy. When oxidized into
the maximal-dimension parent supergravity, the transverse spaces of these domain
walls become cohomogeneity-one spaces with generalized Heisenberg
symmetries and a homothetic conformal symmetry. These metrics may also be
obtained as scaling limits of generalized Eguchi-Hanson metrics, or, with
appropriate discrete identifications, from generalized Atiyah-Hitchin metrics,
thus providing field-theoretic realizations of string-theory orientifolds.
\end{quote}

\section{\large\bf Introduction}

Sets of magnetically charged p-branes with a compact transverse space
necessarily include both positive and negative tension branes. This is because,
although one allows the Bianchi identities for the field strength
$F_{[n]}$ supporting the brane to be violated by magnetic source terms, the
resulting $dF_{[n]}$ should still have a vanishing integral over a compact
space. This ``cohomology condition'' is at the origin of the pairing of positive
and negative tension orbifold planes in the Ho\v{r}ava-Witten construction
\cite{hw} and in the dimensionally reduced interpretation of this construction
in terms of fixed 3-branes in $D=5$ supergravity \cite{losw}. Codimension-one
BPS solutions are characterized by a linear harmonic function, and the equality
of positive and negative charges in such solutions is a consequence of the
necessity for this function to be continuous in the compact transverse space,
thus requiring equality of `upwards' and `downwards' kinks. The same basic
pattern is found in the more phenomenological approach of Randall and Sundrum
\cite{rs} for branes crafted from patched sections of $D=5$ anti de Sitter space.

For purely gravitational solutions, we shall review below how a simple sum-rule
argument establishes the impossibility of resolving all the singularities in
domain-wall spacetimes of this type with a compact transverse space \cite{gkl}.
A domain wall (\ie a $(D-2)$-brane) must be supported by the combined effects of
gravity and a scalar field with a potential. Descending from higher dimensions,
this may be viewed as the result of a dimensional reduction in which the
supporting form field ${\cal F}_{nm}$ has been reduced to a constant 0-form in
the ``internal'' dimensions. The source-free Einstein equations in this case may
be organized so as to have a total derivative on one side and a sum of
non-negative terms on the other. Integration of the total derivative over a
compact transverse space then yields zero, forcing the vanishing of the
derivative of the scalar field that would otherwise have to support the domain
wall, thus ruling out a non-trivial brane spacetime. Of course, if a source is
present the situation is different, and non-trivial spacetimes can in that case
be found. But with a source present, the spacetime has a singularity.

Although it is not possible to find a solution with all
singularities resolved on a compact transverse space, it is still of interest to
look for resolved domain-wall solutions with noncompact transverse spaces. Such a
solution could be viewed as ``half'' of a Ho\v{r}ava-Witten configuration. It is
of particular interest to see whether the negative tension objects can be
resolved, as these are the ones for which no acceptable delta-function source
exists in string theory. Moreover, experience shows that the possibilities for
resolving singularities may be greater if one dimensionally oxidizes a solution
into the highest-dimensional parent supergravity, since dimensional
reduction can introduce singularities in otherwise non-singular spacetimes
\cite{ght}.

The tension $T$ of a brane solution can be read off from the change in the
extrinsic curvature $K_{\mu\nu}$, since by the Israel matching condition for
extrinsic curvatures on $+$ and $-$ sides of a domain wall, one has
\be
\Delta K_{\mu\nu}=K_{\mu\nu}^{(+)}-K_{\mu\nu}^{(-)}=-{8\pi G\over3}T
\label{israel}
\ee
where in coordinates locally adapted to the brane geometry the extrinsic
curvature is given by $K_{\mu\nu}=\ft12
n^{\hat\lambda}\partial_{\hat\lambda}g_{\mu\nu}$, where $n^{\hat\lambda}$ is the
normal to the domain-wall worldvolume. As we shall shortly be considering both
$D$ and $D-1$ dimensional contexts, we adopt the convention that ``hatted''
indices refer to the full embedding spacetime containing the brane, while
``unhatted'' indices refer only to the brane worldvolume coordinates $x^\mu$. In a
BPS domain-wall spacetime characterized by a linear harmonic function $H=c+my$ of
the transverse coordinate $y$, with metric
\be
ds^2 = H^{4\over\Delta(D-2)}dx^\mu dx_\mu + H^{4(D-1)\over\Delta(D-2)}dy^2\ ,
\ee
the extrinsic curvature is proportional to the slope parameter $m$. Thus, in a
standard domain wall setup where one reflects the harmonic function, \eg at
$y=0$, so $H=c+m|y|$, the tension (\ref{israel}) is proportional to $-m$.
Accordingly, for a negative tension domain wall, the warp factor
$e^{2A(y)}=H^{4\over\Delta(D-2)}$ of the embedding $D$-dimensional spacetime {\em
grows} as one moves away from the brane, and conversely, the warp factor for a
positive tension brane {\em decreases} as one moves away from the brane. In the
following, we will review what is known about the possibility of resolving the
singularities at such branes.

\section{Difficulties with Singularity Resolutions}

Quite a lot of effort has been devoted to the question of resolving domain-wall
singularities. This has been done by analysis of the possible renormalization
group flows for the supporting scalar fields \cite{kl,bc} in the embedding
spacetime. Another approach is {\it via} sum rules derived from the Einstein
equations \cite{gkl}. For example, one can start from the action for gravity
coupled to a nonlinear sigma model for a set of scalars $\phi^{\sst I}$, $\st
I=1,\ldots,N$ with a potential $V(\phi)$ and possible sources $J_\alpha$:
\be
I_{g,\phi}=\int
d^{d+1}x\sqrt{-g}\left({1\over2\kappa^2}R-G_{\sst{IJ}}(\phi)
\partial_M\phi^{\sst I}\partial_N\phi^{\sst J}-V(\phi)-\sum_\alpha
J_\alpha(\phi)\delta(y-y_\alpha)\right)\ .
\ee
For a generalized domain-wall spacetime, one takes the metric ansatz
\be
ds^2=e^{2A(y)}g_{\mu\nu}(x)dx^\mu dx^\nu + e^{2B(y)}\ ,
\ee
leading to Einstein equations that can be combined to give on the left-hand side
a total derivative of the warp factor,
\be
A''={\kappa^2\over12}(T_\mu^\mu-4T_5^5)-{1\over12}e^{-2A}R_{\rm brane}\ ,
\ee
where $R_{\rm brane}$ is the Ricci scalar for the brane worldvolume metric
$g_{\mu\nu}$. Substituting for $T_{\mu\nu}$ and $T_{55}$ and integrating 
$\oint dy$ over a compact transverse space, the total derivative $A''$ drops out
and so one has the sum rule
\be
\oint dy\left(g_{\sst{IJ}}\phi^{\sst I}{}'\phi^{\sst
J}{}'+\sum_\alpha J_\alpha(\phi)\delta(y-y_\alpha) +
{1\over4\kappa^2}e^{-2A}R_{\rm brane}\right)=0\ .
\ee
Thus, in order to resolve the singularities of a set of domain walls with a
compact transverse space and with vanishing worldvolume Ricci scalar $R_{\rm
brane}$, the absence of sources $J_\alpha$ would imply $\oint dy\left(
g_{\sst{IJ}}\phi^{\sst I}{}'\phi^{\sst J}{}'\right)=0$ and for positive definite
sigma model metric $g_{\sst{IJ}}$ this would require constant scalar
fields, $\phi^{\sst I}{}'=0\ \forall \st I$. But nontrivial scalars are needed
to support a domain-wall spacetime, so this rules out the possibility of
resolving all the domain-wall singularities in a compact transverse space.

If one cannot resolve a full Ho\v{r}ava-Witten pair of domain walls, then one
can try the next best thing, \ie to resolve just half of it, removing the second
brane to infinity and looking for a nonsingular solution for just one brane.
Here one encounters a further difficulty: ``double-ended'' spacetimes with
growing warp factor as one recedes from the brane cannot be resolved. The best
chance of resolving such a spacetime would be after oxidizing it from $D$
dimensions up into a solution of its maximal-dimension parent
supergravity, \eg into $\hat D=11$, where singularities in dilatonic scalars are
not a problem because there are no such scalars. After this dimensional
oxidation, one is not dealing anymore with a brane of codimension one, but
instead the $(\hat D-D+1)$ dimensional transverse space now becomes a space of
cohomogeneity one, with the $y={\rm const}$ slices representing homogeneous
spaces on which the $(\hat D-D)$ dimensional oxidation has been performed. A
negative tension domain wall in $D$ dimensions with $H=1+m|y|$,
$m>0$, would lift after such dimensional oxidation to a spacetime with a
``saddle,'' reducing to a minimum cross-sectional volume in the extra $d=(\hat
D-D)$ dimensions and then re-expanding. But for solutions to the
purely gravitational sector of $D=11$ supergravity, there cannot be smooth
solutions of this sort, as a result of the Cheeger-Gromoll theorem \cite{gr}, or
as can be seen by the following simple argument \cite{glps}. First transform the
transverse $y$ coordinate into $\tilde y$, the proper distance from the
cross section of minimum volume, and consider the metric ansatz $ds^2=d\tilde
y^2+g_{ij}(x,\tilde y)dx^idx^j$, where $i,j=1,\ldots d=(\hat D-D)$. Let
$V(x,y)=\sqrt{\det g_{ij}}$ and consider $\Theta=V'/V$, which is the expansion
rate of geodesic congruence in this spacetime. This quantity is subject to the
Raychaudhuri inequality
\be
{d\Theta\over d\tilde y} \le -{1\over d}\Theta^2-\Sigma_{ij}\Sigma^{ij}\ ,
\ee
where $\Sigma_{ij}={\partial g_{ij}\over\partial\tilde y}-{1\over
d}g^{rs}{\partial g_{rs}\over\partial\tilde y}g_{ij}$. Thus, if $\Theta$ is
negative at some value of $\tilde y$, it remains negative always, so a
nonsingular ``saddle'' spacetime is not possible, at least in this purely
gravitational setting.

Thus, we now come to consider ``single-sided'' domain walls, whose BPS metric
can be taken to be derived from the harmonic function $H=my$. The ``single-sided'' space is obtained from the double-sided space by
making a $Z_2$ identification between the two halves. {\it A priori,}
the single-sided case may look even less promising for singularity resolution
than the double-sided one, since one is trading in a fairly mild
delta-function singularity in the double-sided metric with $H=c+m|y|$ for a more
serious single-sided singularity in the curvature where $H=0$. However, one has
to remember that problems with singularities can be improved after dimensional
oxidation, so one needs to keep an open mind at this point.

\section{Homothetic Heisenberg Brane Metrics}

Let us now consider a specific class of domain wall solutions \cite{glps} which
oxidize up to a very suggestive class of metrics in $\hat D=11$. As a basic
example, consider a domain wall in $D=8$, supported purely by fields derived
from the $\hat D=11$ metric. In order to generate the scalar potential needed to
support this domain wall in $D=8$, the dimensional reduction down from $\hat
D=11$ needs to have a constant flux turned on in the compact 9 \& 10 directions:
${\cal F}^1_{9\,10}=m$, where ${\cal F}^1$ is the field strength of the
Kaluza-Klein vector ${\cal A}^1$ arising in the initial $\hat D=11
\rightarrow 10$ reduction. The resulting 6-brane metric in $D=8$ is
\be
ds_8^2 = H^{1\over6}dx^\mu dx_\mu + H^{7\over6} dy^2\ .\label{6brane}
\ee
As we have noted above, the best chance for a resolution of the singularity
at the surface where $H=0$ is to be found after oxidation on coordinates
$(z_1,z_2,z_3)$ back up to $\hat D=11$, yielding the metric 
\be
ds_{11}^2=dx^\mu dx_\mu + ds_4^2 \label{oxidized6}
\ee 
where the transverse metric $ds^4$ is
\be
ds_4^2= H^{-1}(dz_1+mz_3dz_2)^2 + H(dy^2+dz_2^2+dz_3^2)\ .\label{k3cand}
\ee

The metric (\ref{k3cand}) has some striking properties \cite{gr}. First of all,
it is K\"ahler. To see this, let us adopt an orthonormal basis of vierbein
1-forms
\bea
e^0 = H^{1\over2}dy\quad &&\quad e^1 = H^{-{1\over2}}(dz_1+mz_3dz_2)\nn\\
e^2 = H^{1\over2}dz_2\quad &&\quad e^3= H^{1\over2}dz_3\ .\label{vierbeins}
\eea
Using these, we can give the K\"ahler form $J=e^0\wedge e^1-e^2\wedge
e^3=\ft12J_{ij}dw^i\wedge dw^j$. where $w^i=(y,z_1,z_2,z_3)$. Using $J_i^j$, one
can make holomorphic $P_i{}^j=\ft12(\delta_i{}^j-\im J_i{}^j)$ and
antiholomorphic $Q_i{}^j=\ft12(\delta_i{}^j+\im J_i{}^j)$ projectors. The Darboux
form for the metric (\ref{k3cand}) is achieved by solving the differential
equations $Q_i{}^j\partial_j\zeta^\mu=0$ for complex coordinates $\zeta^\mu$,
$\mu=1,2$. A sample solution to these equations is 
\bea
\zeta^1&=&z_3+\im z_2\nn\\
\zeta_2&=&y+\im z_1 - \ft14m(z_2^2+z_3^2)+{\im\over2}mz_2z_3\ .
\eea
Changing over to these complex coordinates, the metric becomes
\be
ds_4^2=2g_{\mu\bar\nu}d\zeta^\mu d\bar\zeta^{\bar\nu}=H|d\zeta^1|^2 +
H^{-1}|d\zeta^2+\ft12m\bar\zeta^1d\zeta^1|^2\ ,
\ee
where now
$H=[1+m(\zeta^2+\bar\zeta^2) + \ft12|\zeta^1|^2]^{\ft12}$. The K\"ahler
structure is then made explicit by writing
$g_{\mu\bar\nu}=\partial_\mu\partial_{\bar\nu}K$ with $K=2H^3/(3m^2)$.

Constructing the curvature for the vierbeins (\ref{vierbeins}) shows it to
be self-dual in the $d=4$ Euclidean-signature dimensions $w^i$, and
to have ${\rm SU}(2)$ holonomy. The suspicion arises thus that the metric
(\ref{k3cand}) may have something to do either with the gravitational
instanton K3 or with the Eguchi-Hanson metric. We will confirm these associations
by considering the symmetries of the metric (\ref{k3cand}).

For simplicity, we now let the slope parameter be $m=1$, and rewrite the metric
(\ref{k3cand}) using the proper distance $\tilde y=\ft23 y^{3/2}$ as the radial
coordinate. The metric (\ref{k3cand}) becomes
\be
ds^2=d\tilde y^2 + ({3\tilde y\over2})^{-{2\over3}}\sigma_3^2 + ({3\tilde
y\over2})^{{2\over3}}(\sigma_1^2+\sigma_2^2)\ ,
\ee
where $\sigma_1=dz_2$, $\sigma_2=dz_3$, $\sigma_3=dz_1+z_3dz_2$. The
$\sigma_i$ are 1-forms satisfying the algebra
\be
d\sigma_1=0\qquad d\sigma_2=0\qquad d\sigma_3=-\sigma_1\wedge\sigma_2\
.\label{diffalg}
\ee
They are left-invariant with respect to the {\em Heisenberg group} of $3\times3$
upper-triangular matrices
\be
g=\pmatrix{1&k_2&k_1\cr 0&1&k_3\cr 0&0&1}
\ee
with corresponding Killing vectors
\be
R_1={\partial\over\partial z_2}+z_3{\partial\over\partial z_1}\ ,\qquad
R_2={\partial\over\partial z_3}\ ,\qquad R_3={\partial\over\partial z_1}
\ee
satisfying the Heisenberg algebra 
\be
[R_1,R_3]=[R_2,R_3]=0\ ,\qquad [R_1,R_2]=-R_3\ .
\ee

In addition to the Heisenberg symmetry, the metric (\ref{k3cand}) is
characterized by a rate of growth in volume between $\tilde y_0$ and $\tilde
y$ that goes like $\tilde y^{4\over3}$.

The Heisenberg symmetry and the rate of volume growth for the metric
(\ref{k3cand}) led Gibbons and Rychenkova \cite{gr} to identify it with a
singular (BKTY) limit of the K3 metric identified by Tian \& Yau and Bando \&
Kobayashi \cite{bkty}. In this singular limit, K3 is rendered noncompact: the
parts of the manifold that do not correspond to the metric (\ref{k3cand}) are
``blown away'' to infinity. The Heisenberg (or ``Nil'') symmetry appears only in
this singular limit, since the compact K3 has no Killing vectors. Accordingly, we
shall call such limits ``Heisenberg limits.'' From the present example, we learn
that domain-wall solutions with singularities can sometimes be resolved into
nonsingular spaces after taking a single-sided interpretation of the original
space and then dimensionally oxidizing into the highest-dimension parent
supergravity theory. In this construction, the original single transverse
coordinate $y$ becomes firstly the radial coordinate for a space of cohomogeneity
one, in which the $y={\rm const}$ slices are homogeneous subspaces. In the
subsequent resolution into a smooth space like K3, this cohomogeneity-one
structure may be lost.

Another characteristic property of the metric (\ref{k3cand}), which will
generalize to other examples of this sort, is its homothetic scaling symmetry.
The transformation 
\be
z_1\rightarrow \lambda^2z_1\ ,\qquad z_2\rightarrow \lambda z_2\ ,\qquad
z_3\rightarrow \lambda z_3\ ,\qquad y\rightarrow \lambda y
\ee
causes the metric to scale by a constant factor, $ds_4^2\rightarrow \lambda^3
ds_4^2$. A conformal symmetry of this sort, with a constant scaling factor for
the metric, is known as a ``homothety.'' The corresponding Killing vector is
\be
D=y{\partial\over\partial y}+2z_1{\partial\over\partial
z_1}+z_2{\partial\over\partial z_2}+z_3{\partial\over\partial z_3}\ ;
\ee
the existence of this conformal symmetry derives from the fact that the
Heisenberg algebra is itself scale invariant under
\be
\sigma_1\rightarrow \lambda\sigma_1\ ,\qquad
\sigma_2\rightarrow\lambda\sigma_2\ ,\qquad\sigma_3\rightarrow \lambda^2\sigma_3\
.
\ee
Note in particular that the ``slope'' parameter $m$ is not changed in this
transformation, so, unlike apparently similar integration constants for higher
codimension branes, $m$ does not fix a dimensional quantity that determines how
the geometry changes as one moves away from the $H=0$ surface; indeed, if it did,
there could be no such scaling symmetry. The scaling symmetry is obtained,
however, only for the single-sided metric with
$H=my$ that we are considering here; the double-sided case with $H={\rm
const}+m|y|$ breaks the scaling symmetry at the point of reflection.

\section{Relation to the Eguchi-Hanson metric}

The resolution discussed above for the metric (\ref{k3cand}) is highly implicit,
owing to the fact that the explicit metric on K3 is unknown. At the expense of
retaining the singularity, however, one may find a rather more explicit
relation to a self-dual metric that has frequently been discussed in connection
with approximations to K3, namely the Eguchi-Hanson metric \cite{eh}. The
Eguchi-Hanson metric is
\be
ds^2_{\rm EH} = \left(1+{\tilde Q\over\tilde r^4}\right)^{-1}d\tilde r^2 +\ft14
\tilde r^2\left(1+{\tilde Q\over\tilde r^4}\right)\tilde\sigma_3^2+\ft14\tilde
r^29\sigma_1^2+\sigma_2^2)\ ,
\ee
where the $\tilde\sigma_i$ are left-invariant 1-forms with respect to 
${\rm SU}(2)$, satisfying the algebra
\be
d\tilde\sigma_1=-\tilde\sigma_2\wedge\tilde\sigma_3\ ,\qquad
d\tilde\sigma_2=-\tilde\sigma_3\wedge\tilde\sigma_1\ ,\qquad
d\tilde\sigma_3=-\tilde\sigma_1\wedge\tilde\sigma_2\ .
\ee
If one now performs an In\"on\"u-Wigner contraction,
\be
\tilde\sigma_1=\lambda\sigma_2\ ,\qquad\tilde\sigma_2=\lambda\sigma_2\
,\qquad \tilde\sigma_3=\lambda^2\sigma_3\ ,
\ee
then in the limit $\lambda\rightarrow 0$, one obtains the previous
differential algebra (\ref{diffalg}) for $\sigma_1,\sigma_2,\sigma_3$.

In order to take this limit in the metric, one needs first to rescale
the radial coordinate and the charge, $\tilde r=\lambda^{-1} r$, $\tilde
Q=\lambda^{-6} Q$, and also needs to make a coordinate transformation
$y=\ft14mr^2$. Then, upon making the identification $Q={16\over m^4}$, one
obtains the internal sector (\ref{k3cand}) of the $D=8$ domain wall metric, with
$\sigma_1=mdz_2$, $\sigma_2=mdz_3$, $\sigma_3=m(dz_1+mz_3dz_2)$. Thus the
scaling by $\lambda$ blows up the EH region $\tilde r\approx 0$ into the region
described by the domain-wall metric (\ref{k3cand}).

Note that one is dealing here with a singular version of the Eguchi-Hanson
metric with $\tilde Q>0$ (the non-singular metric has $\tilde Q<0$, and is
complete for $\tilde r\ge|\tilde Q|^{1\over4}$), so this construction does not
actually resolve the domain-wall singularity. 

Within the unscaled EH metric, the far region where $\tilde r\rightarrow\infty$
asymptotically tends to a Ricci-flat cone over RP$^3$. This region has a scaling
symmetry $\tilde r\rightarrow\tilde\lambda\tilde r$, $ds^2_{\rm
EH}\rightarrow\tilde\lambda ds^2_{\rm EH}$, generated by a homothetic
conformal Killing vector $E=\tilde r{\partial\over\partial\tilde r}$. In fact,
one also has in this region $\nabla_\mu E_\nu=g_{\mu\nu}$ and since
$\nabla_{[\mu} E_{\nu]}=0$, it follows that $E$ is a gradient (this is also
clear from its explicit form). So $E$ is also hypersurface orthogonal.

In the middle region of the EH metric, $0<<\tilde r<<\infty$, the $E$ scaling
symmetry is lost. This accords with the fact that the slices at fixed $\tilde r$
are orbits of ${\rm SU}(2)$, which does not have a scaling symmetry.

The region near the EH singularity $\tilde r\approx 0$ blows up, as we have
seen, into the metric (\ref{k3cand}) which has the Heisenberg group symmetry and
also a homothetic conformal Killing vector $D=y{\partial\over\partial
y}+2z_1{\partial\over\partial
z_1}+z_2{\partial\over\partial z_2}+z_3{\partial\over\partial z_3}$. This
conformal Killing vector is not, however, proportional to a gradient, so it is
not hypersurface-orthogonal.

\section{Higher Dimensions and Orientifolds}

Starting from a series of gravitational domain-wall solutions in lower
dimensions, with less preserved supersymmetry than in the above
$D=8\leftrightarrow d=4$ example, one can similarly oxidize up to $\hat D=11$ and
so obtain from their transverse spaces a series of cohomogeneity-one manifolds
of special holonomy with varying amounts of preserved supersymmetry.

For example, we can consider a domain wall in $D=6$ for the theory obtained by
turning on fluxes ${\cal F}_{9,10}$ and ${\cal F}_{7,8}$ for the gravitational
Kaluza-Klein vector ${\cal A}_m$ arising in the original $D=11\rightarrow D=10$
reduction (in the notation of Ref.\ \cite{lpsol}, these fluxes correspond to the
zero-form field strengths ${\cal F}^1_{(0)\,23}$ and ${\cal F}^1_{(0)\,45}$). In
$D=6$, one then obtains a single-sided domain wall metric depending on a linear
harmonic function $H=my$ in the codimension-one transverse space,
\be
ds^2=H^{1\over2}dx^\mu dx_\mu+H^{5\over2}dy^2\qquad \mu=0,1,\ldots,4\ .
\ee
Oxidizing this metric up to $\hat D=11$, one obtains a metric
$d\hat s^2=dx^\mu dx_\mu+ds_6^2$, from which one may extract the $d=6$
cohomogeneity-one part
\be
ds_6^2 = H^{-2}[dz_1+m(z_3dz_2+z_5dz_4)]^2+H^2dy^2+H(dz_2^2+\cdots +dz_5^2)\ .
\label{d6metric}
\ee

Defining an orthonormal basis for this metric by
\bea
e^0=Hdy\ \ \ \,\qquad &&\qquad e^1=H^{-1}[dz_1+m(z_3dz_2+z_5dz_4)]\nn\\
e^2=H^{1\over2}dz_2\ \qquad &&\qquad e^3=H^{1\over2}dz_3\nn\\
e^4=H^{12}dz_4\qquad &&\qquad e^5=H^{1\over2}dz_5\ ,
\eea
one finds a K\"ahler form $J=e^0\wedge e^1-e^2\wedge e^3-e^4\wedge e^5$. The
metric (\ref{d6metric}) has ${\rm SU}(3)$ holonomy, corresponding to
the $\ft14$ supersymmetry preserved by the $D=6$ domain wall.

By a generalization of the arguments in the $d=4$ case presented above,
the $d=6$ metric (\ref{d6metric}) may also be obtained as a Heisenberg limit
of a smooth gravitational instanton, in this case a Calabi-Yau 3-fold
\cite{gr}. Now the volume grows as $\tilde y^{3\over2}$ as one recedes from the
singularity, and, after taking a scaling limit analogous to the BKTY one, the
metric develops a generalized Heisenberg symmetry with respect to which which
the homogeneous $\tilde y={\rm const}$ slices are group orbits having the
structure of a $T^1$ bundle over $T^4$ \cite{glps}. For $H=my$ one finds again a
homothetic conformal Killing vector $D=y{\partial\over\partial
y}+3z_1{\partial\over\partial z_1}+\ft32z_2({\partial\over\partial
z_2}+z_3{\partial\over\partial z_3} +z_4{\partial\over\partial
z_4}+z_5{\partial\over\partial z_5})$. And, similarly to the $d=4$ example, the
$d=6$ metric (\ref{d6metric}) can also be obtained from a
$d=6$ generalization of the Eguchi-Hanson metric \cite{glps}.

Higher-dimensional cohomogeneity-one gravitational domain walls
of special holonomy may also be obtained as Heisenberg limits of
corresponding-dimensional gravitational instantons. For example, one has a
$\ft14$ supersymmetric $D=5\leftrightarrow d=7$ metric with ${\rm G}_2$ holonomy,
\bea
ds_7^2 &=&
H^4dy^2+H^{-2}[dz_1+m(z_4dz_3+z_6dz_5)]^2+H^{-2}[dz_2+m(z_5dz_3-z_6dz_4)]^2\nn\\
&&\hskip 6cm+ H^2(dz_3^2+\cdots+dz_5^2)\ ,
\eea
or a $\ft1{16}$ supersymmetric $D=4\leftrightarrow d=8$ metric with ${\rm
Spin}(7)$ holonomy,
\bea
ds_8^2 &=& H^6dy^2 + H^{-2}[dz_1+m(z_5dz_4+z_7dz_6)]^2 +
H^{-2}[dz_2+m(z_6dz_4-z_7dz_5)]^2\nn\\
&&\qquad + H^{-2}[dz_3+m(z_7dz_4+z_6dz_5)]^2 + H^3(dz_4^2+\cdots +dz_7^2)\ .
\eea

The relations between domain walls and gravitational instantons that we have
reviewed here all have negative tension, as can be seen from their growth in
volume as one recedes from the singularity and in analogy with the
matching condition (\ref{israel}). Moreover, we have seen that there can be
Heisenberg limits to domain wall spacetimes from various previously known
spacetimes: we have seen implicit limits to such domain walls from
nonsingular K3 and Calabi-Yau spaces and also more explicit limits (but without
resolving the singularities) from generalized Eguchi-Hanson spaces. 

Now, in string theory, one does not have acceptable source actions for negative
tension objects, since the corresponding brane waves would have negative kinetic
energy. Thus it is fortunate that the cases we have investigated can actually be
resolved into portions of nonsingular gravitational instanton spaces, which do
not require such sources. The negative tension objects that string theory does
contain, however, are orientifold planes. These avoid having unacceptable
negative energy brane waves since they are fixed in position by virtue of their
$\Z_2$ identifications: the orientifold plane is located at the
$\Z_2$ fixed point. Returning to the effective supergravity theory, one is,
accordingly, interested in finding spacetimes corresponding to orientifold
planes. One such orientifold plane is related to our basic 6-brane example
(\ref{oxidized6},\ref{k3cand}). In this case, the orientifold spacetime
corresponds to the product of $d=7$ Minkowski space times the Atiyah-Hitchin
(AH) metric
\cite{ah}, which can be written in the form
\be
ds^2_{\rm AH} = dt^2 + a^2(t)\sigma_1^2+b^2(t)\sigma_2^2+c^2(t)\sigma_3^2\ .
\ee
At large $t$, one has $a(t)\rightarrow b(t)$ and the AH metric tends
to the Taub-NUT metric, which in appropriate coordinates is
\be
ds^2_{\rm TN} = 4(1+{2M\over r})^{-1}(d\psi+\cos\theta d\phi)+(1+{2M\over
r})(dr^2+r^2(d\theta^2+\sin^2\theta d\phi^2))
\ee
with, however, $M<0$ and which is here taken subject to the $\Z_2$ identification
$(\psi,\theta,\phi)\leftrightarrow (\psi-\pi,\pi-\theta,\phi+\pi)$. Taking a
Heisenberg limit of this metric, one reobtains our conical
spacetime (\ref{k3cand}), but subject now to the $\Z_2$ identification
$(z_1,z_2,z_3)\leftrightarrow (-z_1,-z_2,z_3)$. Accordingly, the AH metric is
described as being ``asymptotically locally conical'' (ALC).

In string theory, the macroscopic curvature of a supergravity solution is
interpreted as arising from the accumulation of charge $\leftrightarrow$ tension
for a stack of microscopic extended objects. Accordingly, one may interpret our
basic 6-brane solution (\ref{oxidized6}) also as the Heisenberg limit of the
metric for a stack of orientifold planes, subject also now to the above $\Z_2$
identification. The ADM mass $M$ of this spacetime is negative, corresponding to
the negative tension. However, there are no negative energy brane waves on such
a background because the $\Z_2$ identification eliminates translational zero
modes. This ALC solution manages to avoid being inconsistent with the positive
mass theorem because its structure at infinity is not simply-connected, owing to
the $\Z_2$ identification, and this interferes with the solution of the Dirac
equation subject to boundary conditions at infinity, which is an essential
element of the positive mass theorem proof. 

Higher-dimensional examples of ALC metrics can be found using the recent explicit
constructions of metrics with Spin(7) holonomy \cite{newspin7} and
$G_2$ holonomy \cite{bggg}.

\end{document}